\documentstyle[epsf,pre,aps]{revtex}
\begin{document}
\twocolumn[\hsize\textwidth\columnwidth\hsize\csname @twocolumnfalse\endcsname
\draft
\title{Theory of Chiral Modulations and Fluctuations in Smectic-A Liquid
Crystals Under an Electric Field}
\author{Jonathan V. Selinger,$^1$ Jianling Xu,$^2$ Robin L. B. Selinger,$^2$
B. R. Ratna,$^1$ and R.~Shashidhar$^1$}
\address{$^1$Center for Bio/Molecular Science and Engineering,
Naval Research Laboratory, Code 6900, \\
4555 Overlook Avenue, SW, Washington, DC  20375 \\
$^2$Physics Department, Catholic University of America, Washington, DC 20064}
\date{January 24, 2000}
\maketitle

\begin{abstract}
Chiral liquid crystals often exhibit periodic modulations in the molecular
director; in particular, thin films of the smectic-C* phase show a chiral
striped texture.  Here, we investigate whether similar chiral modulations can
occur in the {\it induced\/} molecular tilt of the smectic-A phase under an
applied electric field.  Using both continuum elastic theory and lattice
simulations, we find that the state of uniform induced tilt can become unstable
when the system approaches the smectic-A--smectic-C* transition, or when a high
electric field is applied.  Beyond that instability point, the system develops
chiral stripes in the tilt, which induce corresponding ripples in the smectic
layers.  The modulation persists up to an upper critical electric field and
then disappears.  Furthermore, even in the uniform state, the system shows
chiral fluctuations, including both incipient chiral stripes and localized
chiral vortices.  We compare these predictions with observed chiral modulations
and fluctuations in smectic-A liquid crystals.
\end{abstract}

\pacs{PACS numbers:  61.30.Gd, 64.70.Md}

\vskip2pc]
\narrowtext

\section{Introduction}

Molecular chirality leads to the formation of many types of modulated
structures in liquid crystals~\cite{degennes}.  On a molecular length scale,
the fundamental reason for these modulations is that chiral molecules do not
pack parallel to their neighbors, but rather at a slight twist angle with
respect to their neighbors.  On a more macroscopic length scale, molecular
chirality leads to a continuum free energy that favors a finite twist in the
director field.  This favored twist leads to bulk three-dimensional phases with
a uniform twist in the molecular director, such as the cholesteric phase and
the smectic-C* phase.  It also leads to more complex phases with periodic
arrays of defects, such as the twist-grain-boundary phases.

In this paper, we consider the possibility of a new type of chiral modulation
in liquid crystals.  The smectic-A phase of chiral molecules is known to
exhibit the electroclinic effect:  an applied electric field in the smectic
layer plane induces a molecular tilt~\cite{garoff}.  This induced tilt is
generally assumed to be uniform in both magnitude and direction.  Here, we
investigate whether the uniform electroclinic effect can become unstable to the
formation of a chiral modulation within the layer plane.  There are three
motivations for examining this possibility---one theoretical and two
experimental.

\paragraph{Theory:}  The main theoretical motivation for examining this
possibility is that thin films of chiral liquid crystals in the smectic-C*
phase show chiral modulations within the layer plane.  These modulations have
been observed in polarization micrographs of freely suspended
films~\cite{clark}, and have been explained using continuum elastic
theory~\cite{langer,hinshaw1,hinshaw2,jacobs}.  In these modulations, the
molecules form striped patterns of parallel defect walls separating regions
with the favored chiral twist in the molecular director.  Within the narrow
defect walls, the magnitude of the molecular tilt is different from the favored
value in the smectic-C* phase.

Given that in-plane chiral modulations occur in the smectic-C* phase, it is
natural to ask whether analogous modulations can occur in the smectic-A phase
under an applied electric field.  Naively, there are two possible answers to
that question.  First, we might say that the smectic-A phase under an electric
field has the same structure as the smectic-C* phase, because they both have
order in the molecular tilt.  Indeed, under an electric field there is not
necessarily a phase transition between smectic-A and C*~\cite{bahr1,bahr2}.
Thus, we might argue that the smectic-A phase under an electric field should
have in-plane chiral modulations of the form shown in Fig.~1, just as the
smectic-C* phase does.  On the other hand, the applied electric field itself
breaks rotational symmetry in the smectic layer plane, and hence it favors a
particular orientation of the molecular director.  Any modulation in the
director away from that favored orientation costs energy.  For that reason, we
might argue that the smectic-A phase under an electric field should {\it not\/}
have any chiral modulations.  Because these two naive arguments contradict each
other, we must do a more detailed calculation to determine whether chiral
modulations can occur in the smectic-A phase under an electric field.

\paragraph{Experiment 1:}  Apart from these theoretical considerations, striped
modulations have been observed experimentally in the smectic-A phase under an
applied electric field~\cite{crawford1,rappaport}.  The experimental geometry
is shown in Fig.~2(a).  The chiral liquid crystal KN125 is placed in a narrow
cell (2 to 25~$\mu$m wide), and an electric field is applied across the width
of the cell.  One would expect the smectic layers to have a uniform planar
``bookshelf'' alignment, with the layer normal aligned with the rubbing
direction on the front and back surfaces of the cell.  However, the layers
actually form a striped pattern, as shown in Fig~2(b).  In the thicker cells,
the striped pattern is quite complex, with two distinct modulations
superimposed on top of each other~\cite{tang,bartoli}.  The main modulation has
a wavelength of approximately twice the cell thickness, while the higher-order
modulation has a wavelength of approximately 4~$\mu$m, regardless of cell
thickness.  Furthermore, the higher-order modulation is oriented at a skew
angle of approximately 15$^\circ$ with respect to the main modulation, giving
the whole pattern the appearance of a woven texture.

The main modulation in these cells has been explained theoretically as a layer
buckling instability.  When an electric field is applied, the molecules tilt
with respect to the smectic layers, and hence the layer thickness decreases.
Because the system cannot generate additional layers during the experimental
time scale, the layers buckle to fill up space.  This model of layer buckling
predicts layer profiles that are consistent with x-ray scattering from stripes
in 2~$\mu$m cells, which do not have the higher-order modulation~\cite{geer}.
However, this model does not explain the observed higher-order modulation in
thicker cells.  So far, the only proposed explanation of the higher-order
modulation has been a second layer buckling instability in surface regions near
the front and back boundaries of the cell~\cite{bartoli}.  In this paper, we
will consider whether chiral stripes can provide an alternative explanation for
this higher-order modulation.  Such an explanation seems at least initially
plausible because the 15$^\circ$ skew angle between the two observed
modulations suggests a chiral effect.

\paragraph{Experiment 2:}  A separate experimental motivation for considering
chiral modulations in the smectic-A phase comes from measurements of the
circular dichroism (CD), the differential absorption of right- and left-handed
circularly polarized light.  Recent experiments have measured the CD spectrum
of KN125 in the smectic-A phase~\cite{spector}.  When the light propagates
normal to the smectic layers, the CD signal is undetectable.  However, when the
light propagates in the smectic layer plane, in narrow cells as in Fig.~2(a),
the CD signal is much larger, and it is quite sensitive to both electric field
and temperature in a non-monotonic way.  The measured CD signal indicates that
the liquid crystals must have some chiral twist in the layer plane.  This
chiral twist might arise from a bulk phenomenon---either chiral modulations or
chiral fluctuations due to incipient chiral modulations~\cite{lubensky0}.
Alternatively, it might arise from a surface phenomenon, such as the surface
electroclinic effect~\cite{spector}.  Hence, we would like to predict the bulk
chiral modulations and fluctuations in the smectic-A phase as a function of
electric field and temperature, and assess whether such effects can explain the
CD results.

Based on those three motivations, in this paper we propose a theory for chiral
modulations and fluctuations in the smectic-A phase under an applied electric
field.  This theory uses the same type of free energy that has earlier been
used to explain chiral stripes in the smectic-C*
phase~\cite{hinshaw1,hinshaw2,jacobs}, with modifications appropriate for the
smectic-A phase.  We investigate this model using both continuum elastic theory
and lattice Monte Carlo simulations.  In continuum elastic theory, we make an
ansatz for the chiral modulation and insert this ansatz into the free energy
functional.  By minimizing the free energy, we determine whether the induced
molecular tilt is uniform or whether it is modulated in a chiral striped
texture.  In the lattice Monte Carlo simulations, we allow the system to relax
into its ground state, which may be either uniform or modulated, without making
any assumptions about the form of the chiral modulation.  Both calculations
show that the uniform state can become unstable when the temperature decreases
toward the smectic-A--smectic-C* transition, or when a high electric field is
applied. Beyond that instability point, the system develops a chiral modulation
in the molecular tilt, and this tilt modulation induces a corresponding striped
modulation in the shape of the smectic layers.  This modulation is consistent
with the higher-order stripes observed in the smectic-A phase.

In addition to these theoretical results for the chiral modulation, we also
consider chiral fluctuations that occur before the onset of the chiral
modulation itself.  The continuum elastic theory predicts the magnitude of the
incipient chiral stripes as a function of both electric field and temperature.
The lattice Monte Carlo simulations show the incipient chiral stripes as well
as localized chiral vortices in the tilt direction.  These predicted
fluctuations might be visible in future optical experiments.  However, our
predictions for these bulk fluctuations differ from the CD results mentioned
above~\cite{spector} in some important details, including the predominant
direction of the fluctuations and the dependence of the fluctuations on field
and temperature.  Hence, those CD experiments must be showing a chiral surface
phenomenon, such as the surface electroclinic effect.

The plan of this paper is as follows.  In Sec.~II, we propose the free energy
and use continuum elastic theory to predict chiral modulations in the tilt and
the layer shape.  In Sec.~III, we work out the consequences of this theory for
chiral fluctuations in the uniform phase.  In Sec.~IV, we present the lattice
Monte Carlo simulations of chiral modulations and fluctuations in this model.
Finally, in Sec.~V, we discuss the results and compare them with experiments.

\section{Chiral Modulations}

\subsection{Tilt Modulation}

In this theory, we begin by considering a single smectic layer in the $xy$
plane.  Let ${\bf c}(x,y)$ be the tilt director, i.e. the projection of the
three-dimensional molecular director ${\bf\hat{n}}(x,y)$ into the layer plane.
The free energy can then be written as
\begin{eqnarray}
F=\int dA
\biggl[&&\frac{1}{2}r|{\bf c}|^2
+\frac{1}{4}u|{\bf c}|^4
+b{\bf\hat{z}}\cdot{\bf E}\times{\bf c}
-\lambda|{\bf c}|^2{\bf\hat{z}}\cdot{\bf\nabla}\times{\bf c}\nonumber\\
&&+\frac{1}{2}K_S({\bf\nabla}\cdot{\bf c})^2
+\frac{1}{2}K_B({\bf\nabla}\times{\bf c})^2\biggr].
\label{freeenergy1}
\end{eqnarray}
Here, the $r$ and $u$ terms are the standard Ginzburg-Landau expansion of the
free energy in powers of ${\bf c}$.  Near the transition from smectic-A to
smectic-C, we have $r=\alpha(T-T_{AC})$.  The $K_S$ and $K_B$ terms represent
the Frank free energy for splay and bend distortions of the director field,
respectively.  The $b$ and $\lambda$ terms are both chiral terms.  The $b$ term
represents the interaction of the applied electric field ${\bf E}$ with the
molecular director, and the $\lambda$ term gives the favored variation in the
director due to the chirality of the molecules.  (This term is written as
$|{\bf c}|^2 {\bf\hat{z}}\cdot{\bf\nabla}\times{\bf c}$ rather than just
${\bf\hat{z}}\cdot{\bf\nabla}\times{\bf c}$ because the latter term is a total
divergence, which integrates to a constant depending only on the boundary
conditions.  By contrast,
$|{\bf c}|^2 {\bf\hat{z}}\cdot{\bf\nabla}\times{\bf c}$ is not a
total divergence because the factor of $|{\bf c}|^2$ couples variations in the
magnitude of ${\bf c}$ with variations in the orientation.)  This free energy
is identical to the free energy that has been used in studies of smectic-C*
films~\cite{hinshaw1,hinshaw2,jacobs}, except for two small changes.  First, we
now take the coefficient $r$ to be positive, which is appropriate for smectic-A
films without spontaneous tilt order.  Second, we have added the electric field
term, which gives induced tilt order.

We can now ask what configuration of the tilt director ${\bf c}(x,y)$ minimizes
the free energy.  In particular, is the optimum tilt director uniform or
modulated?  To answer this question, we make an ansatz for ${\bf c}(x,y)$ that
can describe both the uniform and modulated states. Suppose the electric field
is in the $y$ direction, ${\bf E}=E{\bf\hat{y}}$, which favors a tilt in the
$x$ direction.  In a uniform state, the system has the electroclinic tilt ${\bf
c}(x,y)=c_0{\bf\hat{x}}$.  By comparison, in a modulated state, the system has
a chiral striped pattern of the form shown in Fig.~1.  Here, the tilt director
is modulated about the average value of $c_0$ in the $x$ direction.  The
magnitude of the tilt is larger when the tilt varies in the direction favored
by molecular chirality, and it is smaller when the tilt varies in the opposite
direction.  This chiral modulation can be represented mathematically by the
ansatz
\begin{mathletters}
\label{ansatz}
\begin{eqnarray}
c_x (x,y) = & c_0 + & c_1 \cos(q_x x + q_y y), \\
c_y (x,y) = &       & c_1 \sin(q_x x + q_y y).
\end{eqnarray}
\end{mathletters}
This ansatz has four variational parameters:  $c_0$ gives the average tilt,
$c_1$ gives the amplitude of the modulation, and $q_x$ and $q_y$ give the
wavevector.  Within this ansatz, $c_1=0$ corresponds to the uniform state, and
$c_1\not=0$ corresponds to the modulated state.  This is not the most general
possible ansatz.  In general, the $c_x$ and $c_y$ components of the modulation
might have different amplitudes, and might not be exactly $90^\circ$ out of
phase.  Furthermore, the modulation might have multiple Fourier components.
These possibilities will be considered in the lattice Monte Carlo simulations
of Sec. IV.  For now, this simple ansatz demonstrates the relevant physics of
the modulation.

To determine whether the optimum tilt director is uniform or modulated, we
insert the ansatz into the free energy and average over position.  The
resulting free energy per unit area can be written as
\begin{eqnarray}
\frac{F}{A}=&&\frac{1}{2}rc_0^2 +\frac{1}{2}rc_1^2
+\frac{1}{4}u c_0^4 +u c_0^2 c_1^2 +\frac{1}{4}u c_1^4
-b E c_0 \nonumber\\
&&-\lambda q_x c_0 c_1^2
+\frac{1}{2}\bar{K}q_x^2 c_1^2 +\frac{1}{2}\bar{K}q_y^2 c_1^2 .
\label{freeenergy2}
\end{eqnarray}
Here, $\bar{K}=\frac{1}{2}(K_S +K_B)$ is the mean Frank constant.  Minimizing
the free energy over $q_x$ and $q_y$ gives
\begin{mathletters}
\label{wavevector}
\begin{eqnarray}
&&q_x=\frac{\lambda c_0}{\bar{K}},\\
&&q_y=0.
\end{eqnarray}
\end{mathletters}
These expressions give the wavevector of the first unstable mode of chiral
modulation.  Note that this wavevector is in the $x$ direction.  This result is
reasonable, because the chiral stripes in the smectic-C* phase also have the
modulation wavevector parallel to the average tilt direction.  Inserting these
expressions back into the free energy gives the result
\begin{eqnarray}
\frac{F}{A}=&&\frac{1}{2}rc_0^2 +\frac{1}{2}rc_1^2
+\frac{1}{4}u c_0^4 +\left(u-\frac{\lambda^2}{2\bar{K}}\right)c_0^2 c_1^2
+\frac{1}{4}u c_1^4 \nonumber\\
&&-b E c_0,
\label{freeenergy3}
\end{eqnarray}
expressed in terms of the two parameters $c_0$ and $c_1$.

Note that this free energy is only thermodynamically stable (bounded from
below) for a certain range of $\lambda$.  The combination of the $u$ and
$\lambda$ terms can be regarded as a quadratic form in the variables $c_0^2$
and $c_1^2$.  The free energy is thermodynamically stable only if this
quadratic form is positive-definite, which requires $\lambda^2<3\bar{K}u$.  If
this condition is not satisfied, then the free energy must be stabilized either
by higher-order terms (such as $|{\bf c}|^6$) not considered here, or by the
constraint $|{\bf c}|\leq 1$, which results from the fact that ${\bf c}$ is the
projection of the three-dimensional molecular director ${\bf n}$ into the layer
plane.

We can now find the minimum of the free energy with respect to $c_0$ and $c_1$.
The extrema of the free energy are given by
\begin{mathletters}
\label{extrema}
\begin{eqnarray}
&&\frac{\partial}{\partial c_0}\frac{F}{A}
=r c_0 +u c_0^3 +2\left(u-\frac{\lambda^2}{2\bar{K}}\right)c_0 c_1^2 -b E =0,\\
&&\frac{\partial}{\partial c_1}\frac{F}{A}
=r c_1 +u c_1^3 +2\left(u-\frac{\lambda^2}{2\bar{K}}\right)c_0^2 c_1 =0.
\end{eqnarray}
\end{mathletters}
One solution of these equations is the uniform electroclinic state, in which
$c_1=0$ and $c_0$ is given implicitly by
\begin{equation}
r c_0 +u c_0^3 -b E =0.
\label{uniform}
\end{equation}
The question now is whether this solution is a minimum or a saddle point of the
free energy.  If it is a minimum, the uniform state is stable; if it is a
saddle point, the uniform state is unstable to the formation of a chiral
modulation.  To answer this question, we calculate the Hessian matrix of second
derivatives at the uniform solution.  The determinant of this matrix is
\begin{equation}
\det\left[\frac{\partial^2}{\partial c_i \partial c_j}
\left(\frac{F}{A}\right)\right]
=\left[r +3 u c_0^2\right]
\left[r +2\left(u-\frac{\lambda^2}{2\bar{K}}\right)c_0^2\right].
\label{determinant}
\end{equation}
If the chiral coefficient $\lambda$ is weak, with $\lambda^2<2\bar{K}u$, then
the determinant is always positive, implying that the uniform state is always
stable.  On the other hand, if $\lambda^2>2\bar{K}u$, then the determinant can
become negative, and the uniform state can become unstable.  In that case, the
system has a second-order transition from the uniform state to the modulated
state when the determinant passes through zero.  This transition can be driven
by increasing the electric field, which increases $c_0$.  It can also be driven
by decreasing the temperature toward the smectic-A--smectic-C* transition,
which reduces $r$ and increases $c_0$.

By combining Eqs.~(\ref{extrema}--\ref{determinant}), we can calculate
properties of the uniform and modulated states around the transition.  First,
the transition occurs at the electric field $E^*$ given by
\begin{equation}
E^*=b^{-1} r^{3/2} u^{-1/2} \left(\frac{\lambda^2}{\bar{K}u}-2\right)^{-3/2}
\left(\frac{\lambda^2}{\bar{K}u}-1\right).
\label{estar}
\end{equation}
On the uniform side of the transition, the tilt is
\begin{equation}
c_0^* = \left(\frac{r}{u}\right)^{1/2}
\left(\frac{\lambda^2}{\bar{K}u}-2\right)^{-1/2}.
\label{c0star}
\end{equation}
On the modulated side of the transition, the amplitude of the chiral modulation
increases as
\begin{equation}
c_1^2=\frac{2b(E-E^*)}{r^{1/2}u^{1/2}}
\frac{\displaystyle\left(\frac{\lambda^2}{\bar{K}u}-2\right)^{3/2}}%
{\displaystyle\left(\frac{\lambda^2}{\bar{K}u}-1\right)
\left(7-2\frac{\lambda^2}{\bar{K}u}\right)}.
\label{c1scaling}
\end{equation}
Furthermore, the wavevector of the chiral modulation is
\begin{equation}
q_x^* = \frac{\lambda c_0^*}{\bar{K}}
=\frac{\lambda}{\bar{K}}\left(\frac{r}{u}\right)^{1/2}
\left(\frac{\lambda^2}{\bar{K}u}-2\right)^{-1/2}
\label{qxstar}
\end{equation}
on the modulated side of the transition.

There is one more constraint on the uniform-modulated transition that we have
not considered yet.  Because the tilt director ${\bf c}$ is the projection of
the three-dimensional molecular director ${\bf n}$ into the layer plane, its
magnitude is limited to $|{\bf c}|\leq 1$.  This constraint implies that there
is an upper critical field $E_{\rm max}$ beyond which the director is locked at
unit magnitude in the $x$ direction.  From Eq.~(\ref{uniform}), the upper
critical field can be estimated as
\begin{equation}
E_{\rm max}\approx b^{-1}(r+u).
\label{emax}
\end{equation}
Above this field, the chiral modulation disappears and the system becomes
uniform again.

By combining the results derived above, we obtain a phase diagram in terms of
the electric field $E$ and the chiral coefficient $\lambda$, which is shown in
Fig.~3.  For large values of $\lambda$, the system has a transition from the
uniform state to the modulated state at the field $E^*$, and then it goes back
into the uniform state at the field $E_{\rm max}$.  At
$\lambda^2\approx\bar{K}(r+2u)$, these two transitions intersect.  Below that
value of $\lambda$, the system remains in the uniform state for all electric
fields, with no modulated state.

\subsection{Layer Modulation}

So far, we have considered only modulations in the tilt director in a
{\it flat\/} smectic layer.  However, a recent theory of the $P_{\beta'}$
rippled phases of lipid membranes shows that any modulation in the tilt
director will induce a modulation in the curvature of the
membrane~\cite{lubensky1,chen1}.  The same theoretical considerations that
apply to modulations in lipid membranes also apply to smectic layers in
thermotropic liquid crystals~\cite{chen2}.  Hence, we can carry over these
theoretical results to predict the curvature modulation induced by the chiral
stripes in the smectic-A phase.  Let $h$ be the height of the smectic layer
above a flat reference plane.  The results of Refs.~\cite{lubensky1,chen1} then
predict
\begin{equation}
\frac{d^2 h}{dx^2}=\frac{\gamma}{\kappa}\frac{d c_x}{dx}
+\frac{\lambda_{\rm HP}}{\kappa}c_x c_y,
\end{equation}
for a modulation in the $x$ direction.  Here, $\kappa$ is the curvature modulus
of the layer, $\gamma$ is the non-chiral coupling between tilt and curvature,
and $\lambda_{\rm HP}$ is the chiral coupling~\cite{helfrich}.  After inserting
the tilt modulation of Eq.~(\ref{ansatz}), with $q_y=0$ as found above, we can
integrate this differential equation to obtain the curvature modulation
\begin{equation}
h(x)=\left(\frac{\gamma c_1}{\kappa q_x}
-\frac{\lambda_{\rm HP} c_0 c_1}{\kappa q_x^2}\right)\sin q_x x
-\frac{\lambda_{\rm HP} c_1^2}{8\kappa q_x^2}\sin 2 q_x x .
\end{equation}
The most important feature to notice about this modulation is that it includes
both $\sin q_x x$ and $\sin 2 q_x x$ terms.  As a result, the layer profile has
the shape shown in Fig.~4(a) (with a highly exaggerated amplitude).  This
modulation has the symmetry $C_2^{(y)}$---it has a two-fold rotational
symmetry, but it does not have a reflection symmetry in the $xy$ or $yz$ plane
because of the chiral coupling $\lambda_{\rm HP}$.

Although our model applies only to a single smectic layer, the symmetry of the
modulation gives some information about the packing of multiple smectic layers.
Because a single layer does not have a reflection symmetry in the $xy$ or $yz$
plane, the packing of multiple layers should not have such a symmetry either.
Instead, the most efficient packing of multiple layers should have the form
shown in Fig.~4(b), with a series of sawtooth-type waves stacked obliquely on
top of each other.  For that reason, this chiral instability should lead to
oblique stripes, which are not parallel to the average layer normal along the
$z$ axis.  The three-dimensional wavevector therefore has a $q_z$ component as
well as a $q_x$ component.  This inclination of the stripes gives a macroscopic
manifestation of the chiral mechanism that generates the stripes.

\section{Chiral Fluctuations}

At this point, let us return to the problem of tilt variations in a single
smectic layer.  In the previous section, we showed that a chiral instability
can give a periodic modulation in the molecular tilt.  Even if the system does
not have a periodic chiral {\it modulation,} it can still have chiral {\it
fluctuations\/} about a uniform ground state, i.e. $c_1$-type fluctuations
about the uniform electroclinic tilt $c_0$.  Such fluctuations can have an
important effect on the optical properties of the system.  For that reason, in
this section we investigate the theoretical predictions for chiral fluctuations
in the uniform phase.

The first issue in this calculation is to determine what quantity gives an
appropriate measure of the strength of chiral fluctuations.  The simplest
possibility is just the chiral contribution to the free energy of
Eq.~(\ref{freeenergy2}),
\begin{equation}
F_{\rm chiral}=-\lambda q_x c_0 c_1^2 .
\end{equation}
An alternative possibility is suggested by a recent model for the transition
between the isotropic phase and the blue phase III~\cite{lubensky2}.  This work
proposed the chiral order parameter
\begin{equation}
\psi={\bf Q}\cdot{\bf\nabla}\times{\bf Q}
=\epsilon_{ijk}Q_{il}\partial_j Q_{kl} ,
\end{equation}
where $Q_{ij}=n_i n_j -\frac{1}{3}\delta_{ij}$ is the tensor representing
nematic order.  In our problem, the three-dimensional nematic director
${\bf\hat{n}}$ can be written as
\begin{equation}
{\bf\hat{n}}\approx {\bf c}+
\left(1-\textstyle{\frac{1}{2}}|{\bf c}|^2\right){\bf\hat{z}},
\end{equation}
and hence the tensor becomes
\begin{equation}
Q_{ij}\approx\left(
\begin{array}{ccc}
c_x^2-\frac{1}{3} & c_x c_y           & c_x \\
c_x c_y           & c_y^2-\frac{1}{3} & c_y \\
c_x               & c_y               & \frac{2}{3}-c_x^2-c_y^2
\end{array}
\right) .
\end{equation}
Into this expression we insert the ansatz of Eq.~(\ref{ansatz}) for the
fluctuations, with $q_y=0$.  After averaging over position, the chiral order
parameter simplifies to
\begin{equation}
\psi=-q_x c_0 c_1^2.
\end{equation}
Remarkably, this result for the chiral order parameter is equivalent to the
expression for $F_{\rm chiral}$, up to a constant factor.  This equivalence
shows that either expression can be used as a measure of the strength of chiral
fluctuations.

We can now calculate the expectation value of the chiral fluctuations.
Applying the equipartition theorem to the free energy of
Eq.~(\ref{freeenergy3}) gives
\begin{equation}
\left\langle\left|F_{\rm chiral}\right|\right\rangle=
\frac{k_B T \lambda^2 c_0^2}{\bar{K}r-(\lambda^2-2u\bar{K})c_0^2} .
\end{equation}
For low fields, we have $c_0=b E/r$.  Furthermore, we have $r=\alpha(T-T_{AC})$
near the smectic-A--smectic-C transition.  Under those circumstances, the
expectation value becomes
\begin{equation}
\left\langle\left|F_{\rm chiral}\right|\right\rangle=
\frac{k_B T \lambda^2 b^2 E^2}
{\bar{K}\alpha^3(T-T_{AC})^3-(\lambda^2-2u\bar{K})b^2 E^2} .
\end{equation}
This expression shows that the system has chiral fluctuations in the uniform
state.  The magnitude of the chiral fluctuations depends on both the applied
electric field and the temperature.  In particular, this measure of the chiral
fluctuations begins at $\langle|F_{\rm chiral}|\rangle=0$ for zero field and
increases as the applied electric field increases.  If $\lambda^2>2\bar{K}u$,
then increasing the field drives the system toward the uniform-modulated
transition, where the magnitude of the fluctuations diverges.  Otherwise,
increasing the field drives the system toward a finite asymptotic value of the
chiral fluctuations.

\section{Monte Carlo Simulations}

To investigate further this model for chiral modulations and fluctuations, we
have done a series of Monte Carlo simulations.  These simulations serve two
main purposes.  First, they allow the system to relax into its ground state,
which may be either uniform or modulated, without the need for any assumptions
about the form of the chiral modulation.  Hence, they provide a test of the
assumed form of the chiral modulation in Eq.~(\ref{ansatz}).  Second, the
simulations provide snapshots of the tilt director field for different values
of the electric field $E$ and the chiral coefficient $\lambda$.  Thus, they
help in the visualization of the chiral modulations and fluctuations.

In the simulations, we represent the tilt director in a single smectic layer by
a discretized $xy$ model on a two-dimensional hexagonal lattice.  Each lattice
site $i$ has a tilt director ${\bf c}_i$ of variable magnitude
$|{\bf c}_i|\leq 1$.  We suppose that $K_S=K_B=\bar{K}$.  The discretized
version of the free energy of Eq.~(\ref{freeenergy1}) then becomes
\begin{eqnarray}
F= &&\sum_i\left[\frac{1}{2}r|{\bf c}_i|^2 +\frac{1}{4}u|{\bf c}_i|^4
+b{\bf\hat{z}}\cdot{\bf E}\times{\bf c}_i\right] \nonumber\\
&&+\frac{2}{3}\sum_{\langle i,j\rangle}\biggl[
-\lambda\left(\frac{|{\bf c}_i|^2+|{\bf c}_j|^2}{2}\right)
{\bf\hat{z}}\cdot{\bf\hat{r}}_{ij}\times({\bf c}_j-{\bf c}_i) \nonumber\\
&&\phantom{+\frac{2}{3}\sum_{\langle i,j\rangle}\biggl[}
+\frac{1}{2}\bar{K}({\bf c}_j-{\bf c}_i)^2\biggr] .
\end{eqnarray}
Here, ${\bf\hat{r}}_{ij}$ is the unit vector between neighboring sites $i$ and
$j$.  The factor of $\frac{2}{3}$ is required because of the hexagonal
coordination of the lattice.  This model is similar to a discretized model for
tilt modulations studied in the context of Langmuir monolayers~\cite{spiral}.

We use a lattice of $100\times100$ sites with periodic boundary conditions.  We
fix the parameters $r=0.5$, $u=1$, $b=1$, and $\bar{K}=1.5$, and vary the
chiral coefficient $\lambda$ and the strength of the electric field $E$ in the
$y$ direction.  For each set of $\lambda$ and $E$, we begin the simulations at
a high temperature, with all the directors ${\bf c}_i=0$, corresponding to an
untilted smectic-A phase.  We allow the system to come to equilibrium and then
slowly reduce the temperature, so that the system can find its ground state.
This procedure can be regarded as a simulated-annealing minimization of the
discretized free energy with a fixed set of parameters.  To refine the phase
diagram further, some simulations of the modulated phase were performed using
system size $100\times4$.  The results are in close agreement with those in the
$100\times100$ system, since the modulation is essentially one-dimensional.

Figure~5 shows the director configuration that forms in a typical run with
$\lambda=2.4$ and $E=0.4$.  Here, the system has relaxed into a configuration
of chiral stripes with wavevector in the $x$ direction, perpendicular to the
electric field applied in the $y$ direction.  This configuration appears
similar to the ansatz proposed in Fig.~1.  The simulation results can be
compared quantitatively with the predictions of continuum elastic theory.  For
these parameters, Eq.~(\ref{estar}) predicts $E^*\approx 0.4$ and
Eq.~(\ref{emax}) predicts $E_{\rm max}=1.5$.  Hence, the system is quite close
to the uniform-modulated transition at $E=E^*$ and far from $E_{\rm max}$.
From Eqs.~(\ref{c0star}) and~(\ref{qxstar}), the average value of the tilt
should be $c_0^*\approx 0.5$ and the modulation wavevector should be
$q_x^*\approx 0.8$, corresponding to a wavelength of $2\pi/q_x^*\approx 8$
lattice units.  These predictions are approximately consistent with the
simulation results.

For a further comparison of the continuum elastic theory with the simulations,
we can look at the shape of $c_y(x)$ as a function of $x$.  The ansatz of
Eq.~(\ref{ansatz}) assumes that this is a sine wave with a single wavevector
$q_x$.  In the simulation, the modulation can take any shape, so we can assess
whether the shape is actually sinusoidal.  Figure~6 shows the shape of the
modulation for two sets of parameters.  In Fig.~6(a) we have $\lambda=2.25$ and
$E=1.48$, which is quite close to $E_{\rm max}\approx 1.5$.  For these
parameters, the amplitude of the modulation is very small, and the shape of the
modulation is very well fit by a sine wave.  By contrast, in Fig.~6(b) we have
$\lambda=2.55$ and $E=0.2$, which is far from the transitions at $E^*$ and
$E_{\rm max}$.  Here, the amplitude of the modulation is much larger, and it is
clearly not sinusoidal.  This comparison shows that the theoretical assumption
of a sinusoidal modulation is appropriate close to the transitions, where the
modulation amplitude is small, but it is not appropriate well inside the
modulated state, where the amplitude is large.

The simulation results for the phase diagram are summarized in Fig.~7.  Note
that this phase diagram has the same structure as the phase diagram predicted
by continuum elastic theory in Fig.~3.  For large $\lambda$, the system goes
from the uniform state to the modulated state at the electric field $E^*$.  The
system remains in the modulated state up to the field $E_{\rm max}$, at which
point it goes back into the uniform state.  As $\lambda$ decreases, the range
of the modulated state in the phase diagram decreases, and it finally vanishes
at $\lambda\approx 2.14$.  For smaller values of $\lambda$, the system stays in
the uniform state for all values of the electric field.  The numerical value of
the phase boundary $E^*(\lambda)$ is shifted somewhat from the theoretical
prediction.  The difference can be attributed to the discretization of the
system, which changes the free energy of the modulated state by a few percent.
This small change in the free energy is enough to give a noticeable shift in
$E^*(\lambda)$.

In addition to these results for the phase diagram in the thermodynamic ground
state, the Monte Carlo simulations also allow us to observe chiral fluctuations
in this model.  As noted above, the simulation procedure involves beginning in
a high-temperature disordered state, allowing the system to come to
equilibrium, and then gradually reducing the temperature to zero.  If we
interrupt the simulations at a nonzero temperature, before the system reaches
the ground state, then we can take a snapshot of the fluctuations.  For
example, Fig.~8 shows a snapshot of the simulations at $\lambda=2.4$ and
$E=0.2$, which has been interrupted at temperature $T=0.081$.  In the
thermodynamic ground state given by the phase diagram, this system is uniform.
At this finite temperature, the uniform state shows chiral fluctuations, which
take the form of incipient chiral stripes.  These incipient chiral stripes show
a specific realization of the chiral fluctuations discussed in Sec.~III.  As
the field is increased, these fluctuations grow in magnitude and eventually
become equilibrium stripes at $E^*(\lambda)$.

In addition to the incipient chiral stripes, the model also shows another type
of chiral fluctuations, which are localized chiral vortices in the tilt
director.  Figure~9 shows an example of the vortices for $\lambda=2.4$ and
$E=0$.  These vortices seem to be forming an incipient hexagonal lattice.  The
vortex pattern is a nonequilibrium fluctuation, which anneals into the uniform
state as the simulation temperature is decreased.  When a small electric field
is applied, the vortices are suppressed.  The field strength required to
suppress vortices increases as $\lambda$ increases.  This nonequilibrium vortex
pattern is similar to the equilibrium hexagonal lattice of vortices that has
been predicted in theoretical studies of the chiral smectic-C*
phase~\cite{hinshaw1,hinshaw2,jacobs}.  The nonequilibrium vortex pattern
probably evolves into the equilibrium vortex lattice as $r$ is decreased from
the smectic-A phase into the smectic-C* phase, but we have not yet tested this
scenario in the simulations.

\section{Discussion}

In the preceding sections, we have answered the theoretical question that
motivated this study.  Our model shows that the smectic-A phase under an
applied electric field can become unstable to the formation of a chiral
modulation within the layer plane, which is similar to the chiral striped
modulation that has been observed in thin films of the smectic-C* phase.  The
transition from the uniform to the modulated state occurs when the temperature
decreases toward $T_{AC}$ or when a high electric field is applied.  It is
somewhat surprising that an applied electric field can induce a modulation in
the director away from the orientation favored by the field.  However, an
analogous effect has been predicted by a recent study of cholesteric liquid
crystals in a field~\cite{seidin}.  That theoretical study showed that a high
electric field can induce a transition from a paranematic phase to a
cholesteric phase, in which the director has a modulation away from the field
direction.  In both that problem and our current problem, the transition
between the uniform and modulated states is controlled by the competition
between field-induced alignment and chiral variations in the orientation of the
field-induced order parameter.

We can now compare this model with the two experimental results mentioned at
the beginning of this paper.  In the first experiment, two types of stripes
were observed in the smectic-A phase of the chiral liquid crystal KN125 under
an applied electric field---a main modulation with a wavelength of
approximately twice the cell thickness and a higher-order modulation with a
wavelength of approximately 4~$\mu$m, independent of cell
thickness~\cite{tang,bartoli}.  The main modulation has been explained as a
layer buckling instability.  Our model of chiral stripes provides a possible
explanation of the higher-order modulation.  In particular, it predicts the
observed symmetry of the modulation---the observed skew angle between the main
stripes and the higher-order stripes in Fig.~2(b) corresponds to the
theoretical packing angle in Fig.~4(b).  Furthermore, it predicts that the
wavelength of the higher-order modulation is determined by material properties
of the liquid crystal, not just by the cell thickness.

One possible objection to this model for the experiment is that the observed
stripe wavelength does not depend on electric field, while the prediction of
Eq.~(\ref{wavevector}a) depends on electric field implicitly through the
uniform tilt $c_0$.  The response to this objection is that the stripe
wavelength is not necessarily in equilibrium.  Rather, the stripe wavelength is
probably determined by the wavelength at the onset of the instability and
cannot change in response to further variations in electric field.  A second
possible objection is that the predicted stripes occur only close to $T_{AC}$,
while the observed stripes occur in the liquid crystal KN125, which does not
have a smectic-A--smectic-C* transition.  The response to that objection is
that KN125 is an unusual liquid crystal with a large electroclinic effect over
a surprisingly wide range of temperature~\cite{crawford2}.  For that reason,
this material can show chiral stripes over a wide range of temperature.  A
further test of the theory would be to see whether the higher-order stripes
occur in a liquid crystal that has the standard temperature-dependent
electroclinic effect near $T_{AC}$, and to see whether these stripes are more
sensitive to temperature.

The second experiment mentioned at the beginning of this paper measured the CD
spectrum of KN125 in the smectic-A phase under an applied electric
field~\cite{spector}.  A very large CD signal was found for light propagating
in the smectic layer plane, much larger than the CD signal for light
propagating normal to the smectic layers.  This large CD signal indicates that
the liquid crystals have some chiral twist in the smectic layer plane.  This
twist might arise from a bulk phenomenon, such as the chiral fluctuations of
the uniform smectic-A phase considered in this paper.  Alternatively, it might
arise from a surface phenomenon, such as the surface electroclinic effect.

Although our model for chiral fluctuations in the uniform smectic-A phase gives
one possible source for a CD signal that could be measured in optical
experiments, our predictions differ from the experimental results in two
important details.  First, the predominant wavevector of the predicted chiral
fluctuations is in the $x$ direction, perpendicular to the electric field.  By
contrast, the experiments are sensitive to chiral fluctuations in the $y$
direction, along the electric field, because the light is propagating in that
direction.  Second, the theory predicts that the quantity
$\langle|F_{\rm chiral}|\rangle$, measuring the strength of chiral
fluctuations, should be zero at zero field and should increase monotonically
with increasing field.  In the experiments, the measured CD signal is nonzero
at zero field, and it can increase or decrease with increasing field, depending
on the temperature.  Thus, the experiment must be showing a chiral surface
phenomenon.  A specific model based on the surface electroclinic effect has
been presented in Ref.~\cite{spector}.  The bulk chiral fluctuations discussed
in this paper might be observable in future optical experiments, especially if
the surface effects can be suppressed through appropriate surface treatments.

In conclusion, we have shown that the uniform electroclinic effect in the
smectic-A phase of chiral liquid crystals can become unstable to the formation
of a chiral modulation in the layer plane.  In the modulated state, there are
stripes in the orientation of the molecular director, analogous to the stripes
that have been observed experimentally in thin films of the smectic-C* phase.
The same mechanism also gives chiral fluctuations in the uniform smectic-A
phase, which grow in magnitude as the modulated state is approached.  These
theoretical results provide a possible explanation for stripes observed in the
smectic-A phase, and they predict chiral fluctuations that may be observed in
future optical experiments.

\acknowledgments

We would like to thank D. W. Allender, A. E. Jacobs, F. C. MacKintosh,
D. Mukamel, R. G. Petschek, and M. S. Spector for helpful discussions.  This
work was supported by the Office of Naval Research and the Naval Research
Laboratory.

\epsfclipon

\begin{figure}
\vbox{\centering\leavevmode\epsfxsize=3.375in\epsfbox{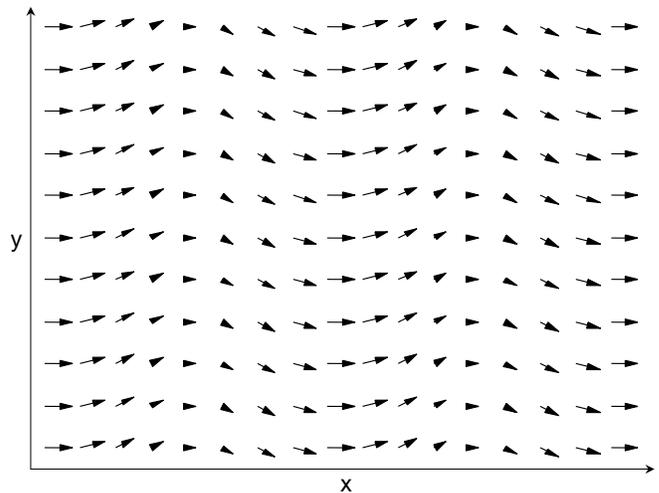}\bigskip
\caption{Ansatz for the chiral modulation of the molecular tilt in a layer of
the smectic-A phase under an applied electric field.  The arrows represent the
projection of the molecular tilt into the layer plane.  The electric field is
applied in the $y$ direction, and the average molecular tilt is in the $x$
direction.}}
\end{figure}

\begin{figure}
\vbox{\centering\leavevmode\epsfxsize=3.375in\epsfbox{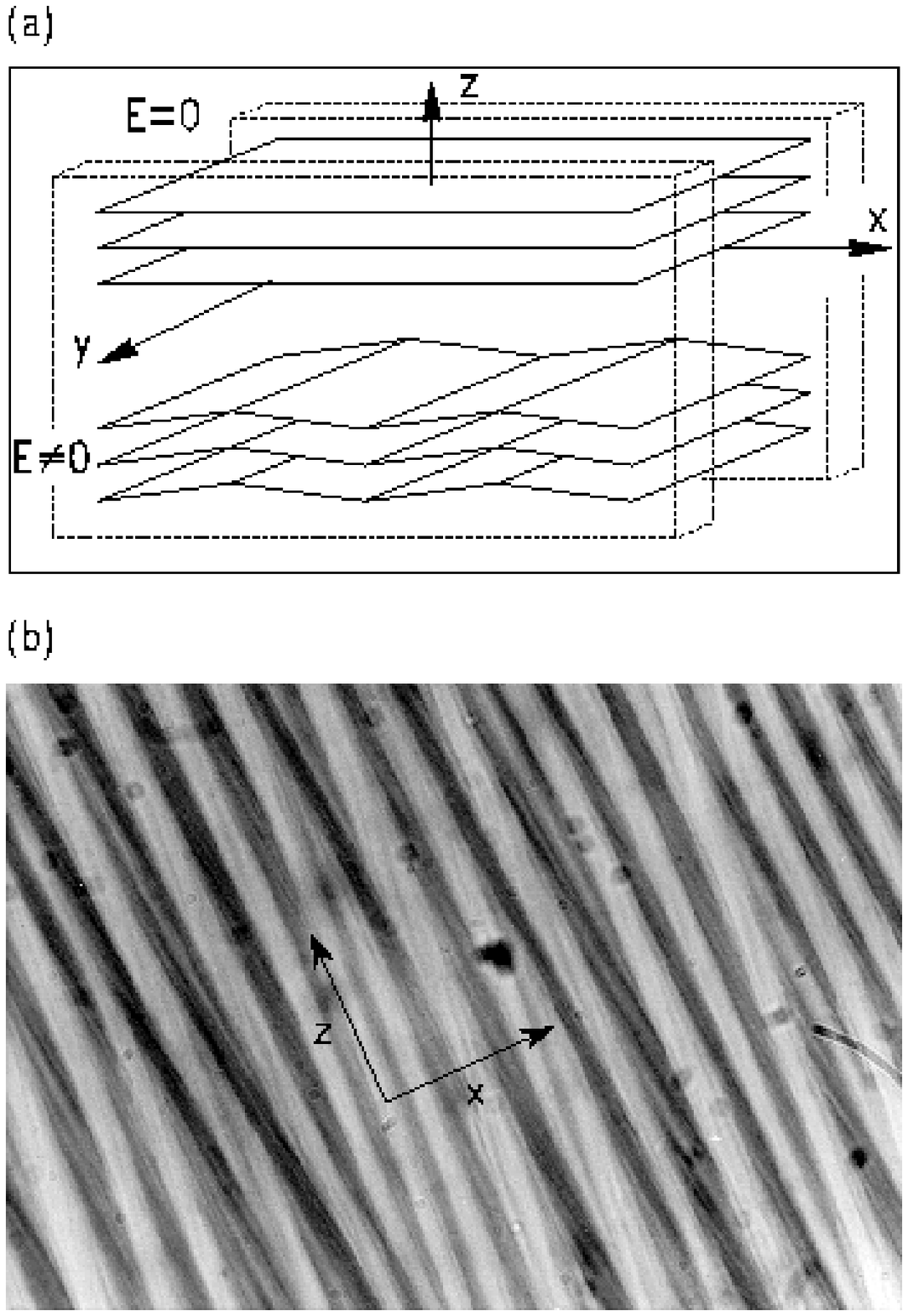}\bigskip
\caption{(a) Experimental geometry of the striped modulation in the smectic-A
phase in narrow cells (from Ref.~\protect\cite{geer}).  (b) Polarization
micrograph of the striped pattern in a 15~$\mu$m cell, showing the $xz$ plane
(with the line of sight along the $y$ axis).  Note that there are two distinct
modulations with wavelengths of approximately 30~$\mu$m and 4~$\mu$m, oriented
at an angle of approximately 15$^\circ$ with respect to each other (from
Ref.~\protect\cite{bartoli}).}}
\end{figure}

\begin{figure}
\vbox{\centering\leavevmode\epsfxsize=3.375in\epsfbox{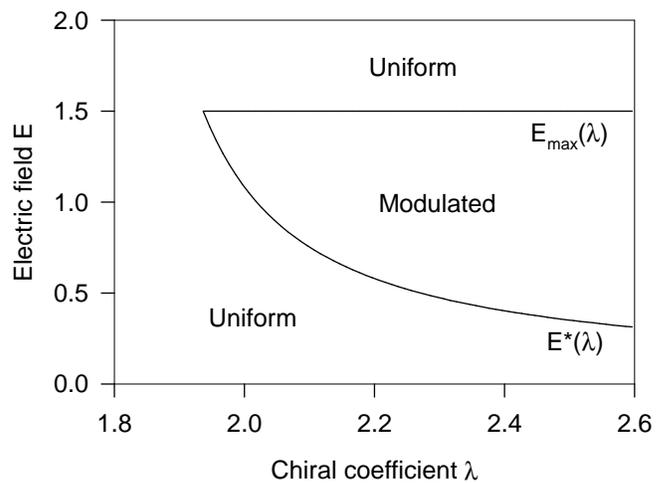}\bigskip
\caption{Theoretical phase diagram in terms of $\lambda$ and $E$ for fixed
$r=0.5$, $u=1$, $b=1$, and $\bar{K}=1.5$, showing the uniform-modulated
transitions at $E^*(\lambda)$ and $E_{\rm max}(\lambda)$.}}
\end{figure}

\begin{figure}
\vbox{\centering\leavevmode\epsfxsize=3.375in\epsfbox{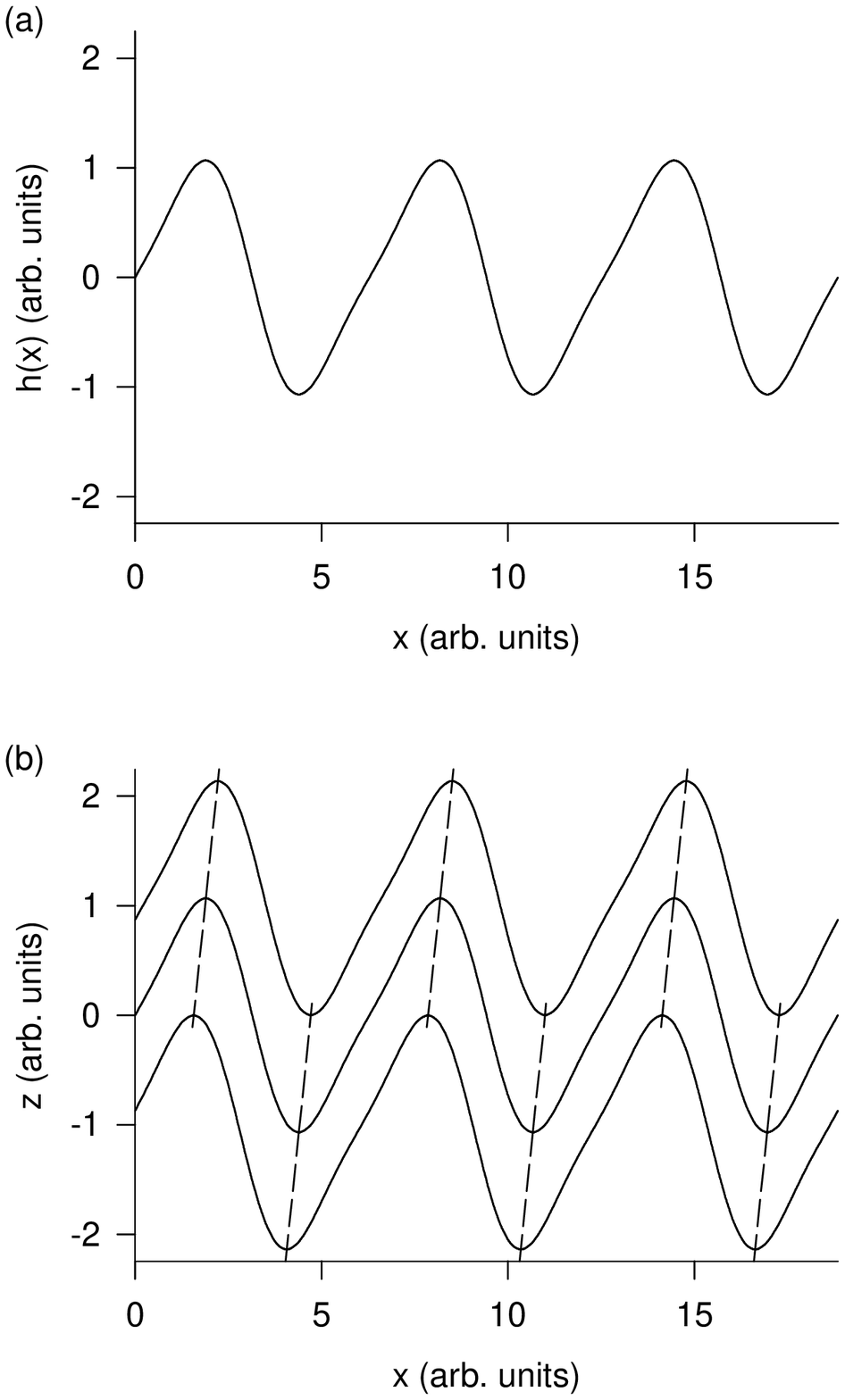}\bigskip
\caption{(a) Profile of the height modulation of a single smectic layer,
showing the $C_2^{(y)}$ symmetry.  The vertical displacement is highly
exaggerated.  (b) The most efficient packing of multiple smectic layers with a
modulation of this symmetry.  The dashed lines indicate the orientation of the
stripes in the $xz$ plane.}}
\end{figure}

\begin{figure}
\vbox{\centering\leavevmode\epsfxsize=3.375in\epsfbox{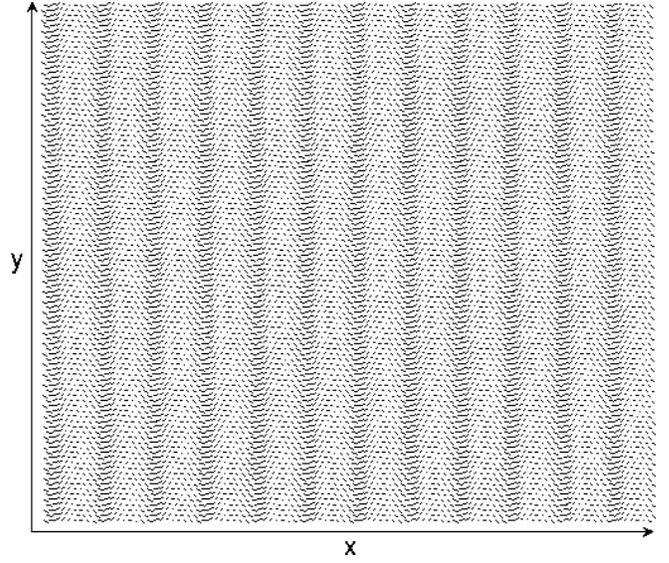}\bigskip
\caption{Simulation results for the molecular tilt in a layer of the smectic-A
phase, showing a chiral modulation for the parameters $r=0.5$, $u=1$, $b=1$,
$\bar{K}=1.5$, $\lambda=2.4$, and $E=0.4$.  The lines represent the projection
of the molecular tilt into the layer plane.  The applied electric field is in
the $y$ direction.}}
\end{figure}

\begin{figure}
\vbox{\centering\leavevmode\epsfxsize=3.375in\epsfbox{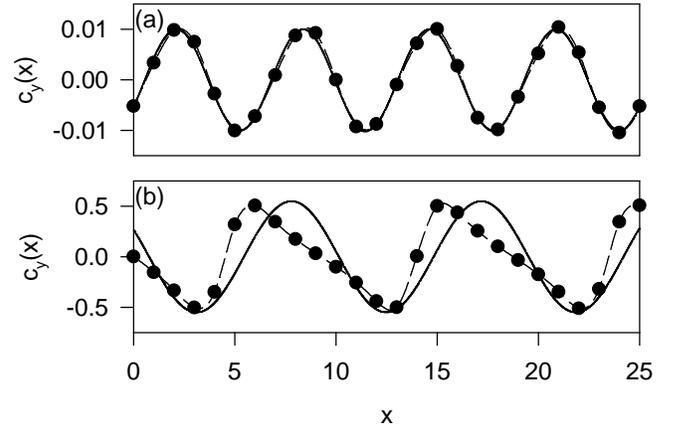}\bigskip
\caption{Simulation results for the shape of $c_y(x)$:  (a) Near a transition,
for $\lambda=2.25$ and $E=1.48$.  (b) Well inside the modulated state, for
$\lambda=2.55$ and $E=0.2$.  Note that the modulation has a sinusoidal form in
the first case, but not in the second.}}
\end{figure}

\begin{figure}
\vbox{\centering\leavevmode\epsfxsize=3.375in\epsfbox{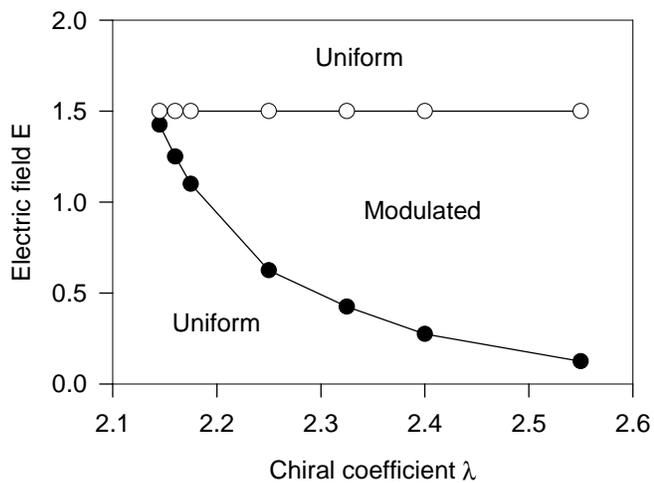}\bigskip
\caption{Simulation results for the phase diagram in terms of $\lambda$ and $E$
for fixed $r=0.5$, $u=1$, $b=1$, and $\bar{K}=1.5$.  The symbols indicate the
observed transitions between the uniform and modulated states.  The discrepancy
between this figure and Fig.~3 is due to the discretization of the lattice.}}
\end{figure}

\begin{figure}
\vbox{\centering\leavevmode\epsfxsize=3.375in\epsfbox{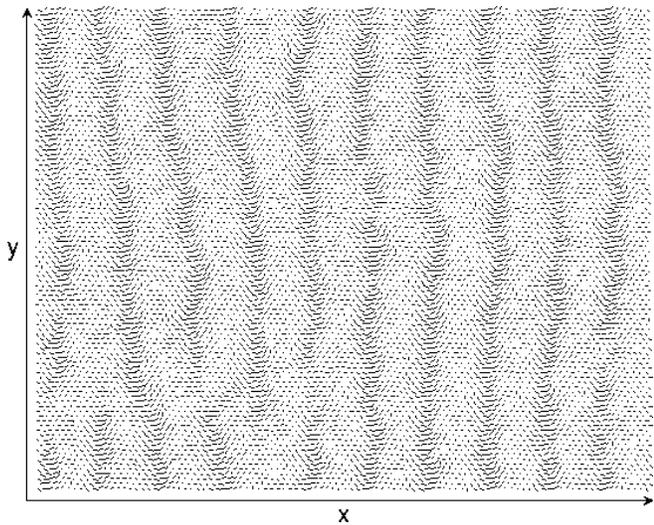}\bigskip
\caption{Finite-temperature simulation results for the parameters $r=0.5$,
$u=1$, $b=1$, $\bar{K}=1.5$, $\lambda=2.4$, and $E=0.2$, showing
incipient chiral stripes.}}
\end{figure}

\begin{figure}
\vbox{\centering\leavevmode\epsfxsize=3.375in\epsfbox{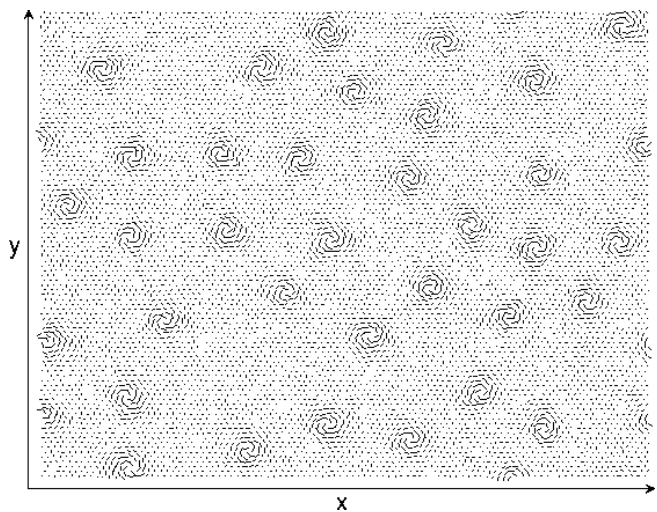}\bigskip
\caption{Finite-temperature simulation results for the parameters $r=0.5$,
$u=1$, $b=1$, $\bar{K}=1.5$, $\lambda=2.4$, and $E=0$, showing localized chiral
vortices.}}
\end{figure}

\end{document}